\documentclass{ws-mpla}
\usepackage[super]{cite}
\usepackage{graphicx}
\begin{document}

\markboth{P. Schulz and G. Wolschin}
{ANALYSIS OF pPb AND PbPb COLLISIONS}

\catchline{}{}{}{}{}

\title{DIFFUSION-MODEL ANALYSIS OF pPb AND PbPb COLLISIONS AT LHC ENERGIES\\
}

\author{P. SCHULZ and {\footnotesize G. WOLSCHIN\footnote{
g.wolschin@thphys.uni-heidelberg.de}}}



\address{ Institut f{\"ur} Theoretische
Physik
der Universit{\"a}t Heidelberg, Philosophenweg 16, D-69120 Heidelberg, Germany, EU}

\maketitle

\pub{Received (Day Month Year)}{Revised (Day Month Year)}

\begin{abstract}
We present an analysis of centrality-dependent pseudorapidity distributions of produced charged hadrons
 in pPb and PbPb collisions at the LHC energy of $\sqrt{s_\text{NN}}=5.02$ TeV, and of minimum-bias pPb collisions at 8.16 TeV
within the nonequilibrium-statistical relativistic diffusion model (RDM). In a three-source approach,
the role of the fragmentation sources is emphasized. Together with the Jacobian transformation
from rapidity to pseudorapidity and the limiting fragmentation 
conjecture, these are essential for modeling the centrality dependence. For central PbPb collisions, a prediction
at the projected FCC energy of $\sqrt{s_\text{NN}}=39$ TeV is made.

\end{abstract}

\ccode{PACS Nos.: 25.75.-q, 24.10.Jv, 24.60.-k}

\section{Introduction}	
In heavy-ion collisions with
energies reached at the Relativistic Heavy-Ion Collider (RHIC) and the Large Hadron Collider (LHC),
one needs to be able to model bulk variables such as the transverse momentum and rapidity or pseudorapidity
distributions of produced charged hadrons. This is an indispensable basis for the detailed understanding of more specific observables 
like quarkonia suppression or jet quenching \cite{wang16}.

To some extent, the equilibrium statistical model for multiple hadron production that was proposed by Fermi \cite{fer50} and Hagedorn \cite{ha68} and later adapted to relativistic heavy-ion collisions by many authors such as Braun-Munzinger et al. or Becattini et al. \cite{mabe08,pbm95,aa06} can be used to model bulk observables, but the approach is more suitable for hadron production rates rather than for distribution functions. The latter are provided by
phenomenological models such as the relativistic diffusion model (RDM), which includes non-equilibrium effects 
to some extent, reproduces substantial features of the data and has predictive power, but does not 
claim to fully account for every detail of the collision and of the ensuing particle production \cite{gw99,gw13}.

The relativistic diffusion model is thus in scope and character located between the statistical model for multiple hadron production, and much more detailed numerical models that aim at a microscopic description of the collision, such as the Color Glass Condensate (CGC, see Ref.\cite{ge10}) for the initial state, hydrodynamics for the main part of the time evolution (e.g. \cite{koi07,luro08,alv10,hesne13}), and codes like URQMD for the final state \cite{bas13}. 

Many other models that account for multiple hadron production in relativistic collisions are available in the literature. For pp, a three-source model with a central fireball accounting for annihilation contributions had been developed in Ref.\cite{inno79} and was also extended to $e^+e^-$, but not to heavy-ion collisions. A three-fireball model for non-diffractive hadron-hadron collisions was published in Ref.\cite{meng83}\,. Relevant for relativistic heavy-ion collisions that are treated in this work are, in particular, three-fluid hydrodynamic calculations such as those in Refs. \cite{ton06,iva16}\,, but these are so far confined to energies accessible at the Relativistic Heavy Ion Collider, they have not yet been performed at LHC energies. They assume two counter-streaming fluids corresponding to the constituent nucleons of the incoming collision partners as in Refs. \cite{ams78,strott86}\,, and a third source (fireball) in the midrapidity region that is associated with a fluid which is net-baryon free.  

Our application of a three-source model in the relativistic diffusion approach to (pseudo-)rapidity distributions of produced charged hadrons incorporates new aspects which were not relevant for the early works about charged-hadron production in hadronic collisions. In particular, the central fireball source is absent in stopping 
(p minus $\bar{\text{p}}$) because particles and antiparticles are produced in equal amounts in low-$x$ gluon-gluon collisions. Moreover, the central source was found to be insignificant in heavy-ion collisions at cm energies below $\approx 20$ GeV per nucleon pair \cite{kw07}. Here, the two fragmentation sources alone account for the produced charged hadrons, because there is no significant density of thermalized low-$x$ gluons available in the system.

A directly related phenomenological model is the one by Liu et al.\cite{liu11}\,, who had conceived a multisource thermal model with originally four sources that was resonably consistent with charged-hadron production in pp, p$\bar{\text{p}}$ and heavy-ion collisions according to $\chi^2$-fits, but lacked a midrapidity source. Following the concept of the three-source relativistic diffusion model \cite{gw99,gw13} \,, Liu's thermal model \cite{liu11} was later modified \cite{liu15} to contain a central and two fragmentation sources, resulting in a clear physical picture with substantially better fits, in particular for produced charged hadrons, but also for stopping data and quarkonia production in relativistic heavy-ion collisions.

In this work we make use of the relativistic diffusion model to analyze centrality-dependent charged-hadron production in pPb and PbPb collisions at the LHC energy of 
$\sqrt{s_\text{NN}}=5.02$ TeV, and in minimum bias  pPb collisions at 8.16 TeV. The model parameters are determined in $\chi^2$-minimizations with respect to the corresponding ALICE and CMS data. 
In the next chapter, we review some essential ingredients of the relativistic diffusion model following Ref. \cite{gw13}\, and referring to previous publications for details. In section 3, we discuss minimum-bias pPb collisions
at 5.02 TeV and 8.16 TeV, and centrality-dependent results  at 5.02 TeV. In section 4, centrality-dependent results for PbPb at 5.02 TeV are investigated. A prediction for central collisions at the projected FCC energy of 39 TeV is made in section 5. The conclusions are drawn in section 6.
\section{Relativistic diffusion model}
\subsection{Fokker-Planck equation with three sources}
In  the relativistic diffusion model with three sources for charged-hadron production, rapidity distributions of produced particles
are calculated from an incoherent superposition of the fragmentation sources $R_{1,2}(y,t=\tau_\text{f})$ with charged-particle content $N_\text{ch}^{1}$, $N_\text{ch}^{2}$ and
the midrapidity gluon-gluon source $R_\text{gg}(y,t = \tau_\text{f})$ at the freeze-out time $t=\tau_\text{f}$ with charged-particle content $N_\text{ch}^\text{gg}$. In the present work, we follow the current convention \cite{alb17} of the LHC-collaborations 
that the forward direction $(R_1)$ in pPb with positive rapidity is defined as p-going, whereas the backward direction $(R_2)$ is Pb-going. The convention was opposite in our analysis \cite{sgw15} of
the initial (2012) Pbp pilot run \cite{ab12}\,.

The three-source pseudorapidity distribution of produced charged hadrons then reads

\begin{equation}
\frac{dN_\text{ch}(y,t=\tau_\text{f})}{dy}=N_\text{ch}^{1}R_{1}(y,\tau_\text{f})
 +N_\text{ch}^{2}R_{2}(y,\tau_\text{f})
+N_\text{ch}^\text{gg}R_\text{gg}(y,\tau_\text{f})\,.
\label{normloc1}
\end{equation}
The rapidity is defined as $y = 0.5\cdot \ln((E+p)/(E-p))$, and the freeze-out time $\tau_\text{f}$ corresponds to the total integration time of the underlying partial differential equation. It is also called interaction time $\tau_\text{int}$ in related publications.

In the linear version of the RDM \cite{gw99}, the macroscopic distribution functions are solutions of a Fokker-Planck equation (FPE)  with a relaxation ansatz for the drift towards
the equilibrium value $y_\text{eq}$, and $k = 1, 2, 3$
\vspace{.2cm}
\begin{equation}
\frac{\partial}{\partial t} R_{k}(y,t) =
-\frac{1}{\tau_{y}^k}\frac{\partial}
{\partial y}\Bigl[(y_\text{eq}-y)\cdot R_{k}(y,t)\Bigr]
+D_{y}^{k} \frac{\partial^2}{\partial y^2}
R_{k}(y,t)\,.
\label{fpe}
\end{equation}
\vspace{.2cm}

The use of the additive variable rapidity -- rather than longitudinal momentum -- in the nonequilibrium-statistical 
Fokker-Planck framework has proven to be a useful approach in calculations and predictions of macroscopic distribution functions for produced particles. The three subdistributions add up incoherently to produce the observed distribution function at $t=\tau_\text{f}$. For $t\rightarrow \infty$, the stationary distribution is reached, which in case of the linear model is simply a Gaussian centered at $y_\text{eq}$, or the corresponding value in the laboratory system for asymmetric collisions.

Hence, in the linear model that we summarize in this review section, the stationary solution actually differs slightly from the thermal equilibrium result that is given by the Boltzmann distribution ($E=m_\perp \cosh(y)$)
\begin{equation}
\label{boltzmann}
  E \frac{d^3N_\text{ch}}{dp^3} \Bigr|_\text{eq}\propto E \exp\left(-E/T\right)
\end{equation}
with the equilibrium temperature $T$.
The deviation from the thermal result is due to the relaxation ansatz in the FPE. To enforce the Boltzmann distribution as an exact equilibrium solution of the model \cite{fgw17}, the drift term must assume the nonlinear form $A\sinh(y)$ with an amplitude given by the temperature $T$, the diffusion coefficient $D$, and the transverse mass $m_\perp = \sqrt{m^2+ p_\perp^2}$ ~($p_\perp$ the transverse momentum) as 
\begin{equation}
A=\frac{m_\perp D}{T}\,,
\label{fdt}
\end{equation}
which is a special expression \cite{asgw18} of the so-called fluctuation-dissipation relation that connects drift and diffusion.

With such a nonlinear drift term, the transport equation can only be solved numerically. Since extensive comparison with
data has shown \cite{gw13} that the analytical linear model already provides physical insights and predictions, we pursue it in this work, and use it for the interpretation of symmetric and asymmetric collisions at LHC energies.

Integrating the FPE with the initial conditions
$R_{1,2}(y,t=0)=\delta(y\mp y_\text{max})$, the absolute value of the beam rapidities 
$y_\text{max}$, and $R_{3=\text{gg}}(y,t=0)=\delta(y-y_\text{eq})$  
yields the exact solution. The mean values 
are derived analytically from the moments 
equations in the center-of-mass system (cms) as
\begin{equation}
<y_{1,2}(t)>=y_\text{eq}[1-\exp(-t/\tau_{y}^{1,2})] \pm y_\text{max}\exp{(-t/\tau_{y}^{1,2})}
\label{mean}
\end{equation}
for the sources (1) and (2) with the absolute value of the beam rapidity $y_\text{max}$ and the rapidity relaxation time $\tau_y$.

The local equilibrium value $y_\text{eq}$ is equal to zero for
symmetric systems, but for asymmetric systems such as pPb, the midrapidity source is moving
\cite{wob06}, and the superposition of the sources is more sensitive to
the values of the model parameters than in the symmetric case. From energy-momentum
conservation, the centrality-dependent equilibrium value in the cms is obtained as \cite{bha53,nag84,wobi06}
\begin{equation}
y_\text{eq}(b)=-0.5\cdot\ln{\frac{\langle m^1_\perp(b)\rangle\exp(-y_\text{max})+\langle m^2_\perp(b)\rangle\exp(y_\text{max})}
{\langle m^2_\perp(b)\rangle\exp(-y_\text{max})+\langle m^1_\perp(b)\rangle\exp(y_\text{max})}}
\label{eq}
\end{equation}
with the beam rapidities $y_\text{beam}^{1,2}=\pm y_\text{max}=\pm \ln({\sqrt{s_\text{NN}}/m_p)}$, the average transverse masses $\langle m^{1,2}_\perp(b)\rangle =
\sqrt{m^2_{1,2}(b)+\langle p_\perp^{1,2}\rangle^2}$, and participant masses $m_{1,2}(b)$ of the p- and Pb-like participants in pPb collisions that depend on the impact parameter $b$. The minus sign refers to cases where $m^2_\perp >m^1_\perp$ such as in the ALICE pPb experiments of 2013 and 2016 where the p beam defined the positive rapidity (forward, 1-direction). The sign of the equilibrium value in the center-of-mass system changes when the beams are interchanged, as was done in the experiments to cover the full phase space.
For sufficiently large beam rapidities $y_\text{max}$ such at LHC energies, the equilibrium value can be approximated as
\begin{equation}
y_\text{eq}(b)\simeq -0.5\cdot\ln{\frac{\langle m^2_\perp(b)\rangle}{\langle m^1_\perp(b)\rangle}}\,.
\hspace{.2cm}
\label{eq1}
\end{equation}

The corresponding numbers of participants can be obtained from the geometrical overlap, or from Glauber calculations. The time evolution in the RDM causes a drift of the distribution functions $R_{1,2}$ towards $y_\text{eq}$. For $y_\text{beam}=\pm8.586$ in the center-of-mass system as in 5.02 TeV pPb an estimate in minimum-bias collisions with $<N_\text{part}>=7.87$\, \cite{adam15} is $y_\text{eq} \simeq -0.946$, and smaller absolute values for more peripheral collisions. In very peripheral collisions with $m^2_\perp(b)\sim m^1_\perp(b)$, the equilibrium value in the cms becomes approximately zero.

Whether the mean values of $R_1$ and $R_2$ actually attain $y_\text{eq}$ depends on the centrality-dependent freeze-out time $\tau_\text{f}$ (the time the system interacts strongly, corresponding to the integration time of Eq.\,(\ref{fpe})), and its ratio to the rapidity relaxation time $\tau_y$. Typical freeze-out times at LHC energies from dynamical models in central
PbPb collisions are 6-8 fm/$c$, which is too short for the fragmentation sources to reach equilibrium, such that their mean values $<y_{1,2}>$ remain between beam and equilibrium values.  

The third source $R_\text{gg}$ already emerges near equilibrium at the parton formation time
and spreads in time due to strong diffusive interactions with other particles, without any shift in the mean value for a given centrality class.
The variances are $(k=1,2,3)$
\begin{equation}
\sigma_{k}^{2}(t)=D_{y}^{k}\tau_{y}^k[1-\exp(-2t/\tau_{y}^k)]\,,
\label{var}
\end{equation}
they reach equilibrium faster than the mean values. Here the diffusion coefficients in rapidity space are $D_y^k$, and presently we assume equal values for the three sources, whereas the relaxation times $\tau_{y}^k$ may differ, causing different widths for the three subdistributions. 

The sources remain clearly separated, although not directly visible in the data for produced charged hadrons. The fragmentation sources do appear in net proton (proton minus antiproton) data where the midrapidity source cancels out, but net-proton distributions are not available at the LHC in a sufficiently large rapidity range.

A microscopic model for the calculation of $\tau_y$ and $D_y$ is not yet available. When comparing to data, it is therefore convenient to use instead the mean values of the fragmentation sources $<y_{1,2}(t=\tau_\text{f})>$ and the variances $\sigma_{k}^{2}(t=\tau_\text{f})$, or FWHMs 
$\Gamma_k$ as parameters.

In case of asymmetric systems, one has, moreover, to consider the fact that in the laboratory system, the center-of-mass moves with a rapidity of
\begin{equation}
\Delta y=\frac{1}{2}\ln\left(\frac{Z_1A_2}{Z_2A_1}\right)
\end{equation}
in the direction of the light (1) beam. For pPb, this shifts the cms by $\Delta y=0.465$ towards forward (positive) rapidities in the laboratory  (for dAu, $\Delta y=0.110$). As a result, the equilibrium value of the rapidity Eq.\,(\ref{eq})
is reduced accordingly when the results of the calculations are shifted to the laboratory system in order to compare with data. In very peripheral collisions with small numbers of participants, the equilibrium value of the rapidity is close to zero, but the shift of the cms relative to the laboratory system must always be taken into account.

To reduce the number of RDM-parameters for asymmetric systems, we had proposed in Ref.\cite {wobi06} to relate the number of produced charged hadrons in the fragmentation sources with the corresponding number of participants
\begin{equation}
N^{1,2}_\text{ch}=N^{1,2}_\text{part}\frac{N^\text{tot}_\text{ch}-N^\text{eq}_\text{ch}}{N_\text{part}^{1}+N_\text{part}^{2}}\,.
\label{nch}
\end{equation}
With this conjecture the model has seven parameters for asymmetric system. For symmetric systems like PbPb there are five parameters.
\subsection{Jacobian transformation and limiting fragmentation}
The LHC data for charged-hadron production are available in pseudorapidity space.
Since the theoretical model is formulated in rapidity space, one has to transform the calculated distribution functions to pseudorapidity space, $\eta=-$ln[tan($\theta / 2)]$, in order to be able to compare with the data, and perform $\chi^2-$minimizations. The Jacobian transformation\\ 
\begin{equation}
\frac{dN}{d\eta}=\frac{dN}{dy}\frac{dy}{d\eta}=
J(\eta,  m / p_\perp)\frac{dN}{dy}, 
\label{deta}
\end{equation}
\begin{equation}
{J(\eta,m /p_\perp)=\cosh({\eta})\cdot }
[1+(m/ p_\perp)^{2}
+\sinh^{2}(\eta)]^{-1/2}
\label{jac}
\end{equation}
depends on the squared ratio of the mass and the transverse momentum of the produced particles.
Hence, its effect increases with the mass of the particles, and it is most pronounced at small transverse momenta. For reliable results one has to consider the full $p_\perp-$distribution, however: It is not sufficient to consider only the mean transverse momentum $\langle p_\perp \rangle$. In Ref.\cite {rgw12} we have discussed how this can be done for known $p_\perp-$distributions of identified $\pi^-, K^-$, and antiprotons. Since the Jacobian depends only on the ratio $m/p_\perp$, we use the pion mass $m_{\pi}$ rather than the mean mass $<m>$, and calculate an effective mean transverse momentum 
$<p_\perp^\text{eff}>$ such that the experimentally determined Jacobian $J_{y=0}$ of the charged-hadron distribution for $\pi^-, K^-$ and $\bar{p}$ at rapidity  zero is exactly reproduced. This yields
for a given centrality class \cite{rgw12}
\begin{equation}
\langle p_\perp^\text{eff}\rangle=m_{\pi}J_{y=0}\Bigr{/}\sqrt{1-J_{y=0}^2}\hspace{.2cm} .
\end{equation}
These effective transverse momenta are smaller than the mean transverse momenta determined from the $p_\perp-$distributions, and the corresponding effect of the Jacobian is therefore larger than that estimated
with $\langle p_\perp \rangle$ taken from the transverse momentum distributions for each particle species. The parameter that defines the Jacobian is then $q=m_\pi/\langle p_\perp^\text{eff}\rangle$.
At high RHIC and LHC energies the effect of the Jacobian transformation remains essentially confined to the midrapidity source.

The Jacobians can now be calculated for each centrality class, pseudorapidity distributions of produced charged hadrons are obtained in the three-source model from Eq.\,(\ref{normloc1}), the parameters are optimized with respect to the available data, and conclusions regarding the relative sizes of the sources become possible.

However, LHC data are still missing in the fragmentation region, and $\chi^2$-optimizations with respect to data in the midrapidity region become rather uncertain in the tails of the distributions where the fragmentation sources become important. We have therefore proposed in Ref.\cite {rgw12} to use the well-known limiting fragmentation, or extended longitudinal scaling effect \cite{ben69,bb03} as an additional constraint: At sufficiently high energy, particle production in the fragmentation region becomes almost independent of the collision energy.

 In particular, in Ref.\cite {rgw12} we used 0.2 TeV AuAu results at RHIC -- where data in the fragmentation region are available \cite{bea02,alv11} -- to supplement the LHC 2.76 TeV PbPb data in analogous centrality classes at larger values of pseudorapidity, shifting the latter by $\Delta y= y_\text{beam}^\text{LHC}- y_\text{beam}^\text{RHIC}=7.99-5.36=2.63$. The resulting RDM-parameters had physically reasonable dependencies on the cm energy and centrality.

It should be noted that whereas limiting fragmentation scaling has been firmly established experimentally in AuAu collisions at RHIC cm energies of 19.6-200 GeV \cite{bea02,bb03,alv09,alv11},
its validity is still debated 
at LHC energies: Due to the lack of data in the fragmentation region one has to rely on phenomenological models such as the
thermal model -- which predicts a violation of limiting fragmentation at LHC energies \cite{cley08} -- or microscopic models, for example, the multiphase transport model AMPT by Ko et al.\cite{ko05,basu16}. 

However, the ALICE collaboration has found in Ref.\cite{abb13} -- in agreement with our results in Ref.\cite{rgw12} -- that their 2.76 TeV PbPb data are (with respect to the 62.4 GeV and 200 GeV AuAu data)
consistent with the validity of extended longitudinal scaling within the errors which arise mainly from the extrapolation of the charged-particle pseudorapidity density from the measured region to the rapidity region of the projectile. There the extrapolation is based on a double-gaussian fit of the data in $\eta$-space. 

In this work the emphasis is on LHC PbPb data at the higher center-of mass energy of 5.02 TeV, which currently cover an even smaller pseudorapidity range than at 2.76 TeV.
Again we use the limiting-fragmentation conjecture, supplementing the 5.02 TeV data in the tail region with 200 GeV data points that are shifted by the difference in beam rapidities, 
$\Delta y= y_\text{beam}^\text{5.02\,TeV}- y_\text{beam}^\text{200\,GeV}=8.59-5.36=3.23$, and with four 2.76 TeV points shifted by $\Delta y=0.6$. The results will be discussed in section\,4.

For pPb collisions, no data at RHIC energies exist, so that limiting fragmentation scaling can not be used to assess the fragmentation region at LHC energies. The 
available data at 5.02 TeV are presently confined\cite{ab12} to the range $|\eta | \lesssim 2$.
Forthcoming LHC pPb data in the pseudorapidity region $2<|\eta|<5$ will then provide more definite conclusions regarding the relative sizes of fragmentation and midrapidity sources.
In the next section, we compare the three-source RDM results over the full pseudorapidty range with the available minimum-bias and centrality-dependent data in the midrapidity region, and determine the RDM parameters in $\chi^2$-minimizations.
\begin{figure}[tph]
\begin{center}
\includegraphics[width=12cm]{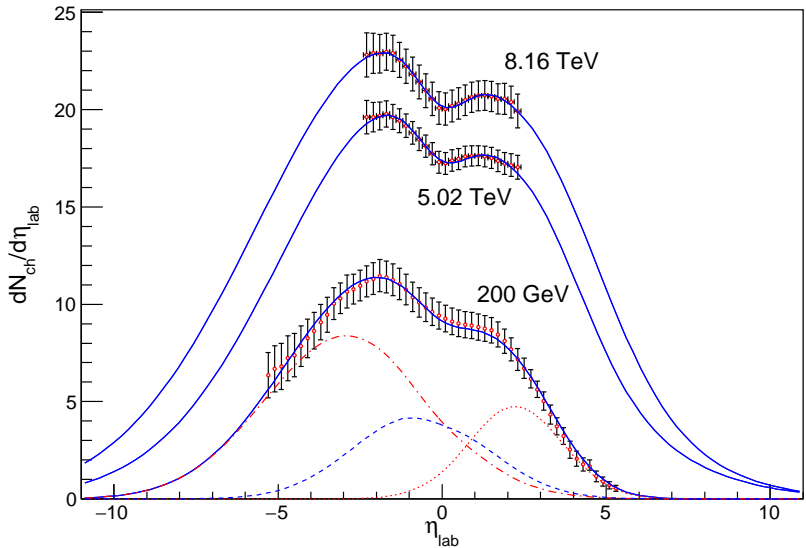}
\caption{\label{fig1}(Color online) The RDM pseudorapidity density distribution functions for produced charged hadrons in minimum-bias pPb collisions
at the LHC cm energy of 8.16 TeV (upper solid curve) is shown together with the result at 5.02 TeV (middle curve, see also Ref.\cite{gw13}) in fits to the CMS data \cite{cms18}{}. The lower curve is the minimum-bias distribution function in dAu at 0.2 TeV measured in Ref.\cite {bb04} with the RDM result in analogy to Ref.\cite {wobi06} and the distribution functions for the three sources (dotted p-going, dashed central, dot-dashed Pb-going). Parameters are from table\,\ref{tab1}. } 
\end{center}
\end{figure}
\begin{figure}[tph]
\begin{center}
\includegraphics[width=12cm]{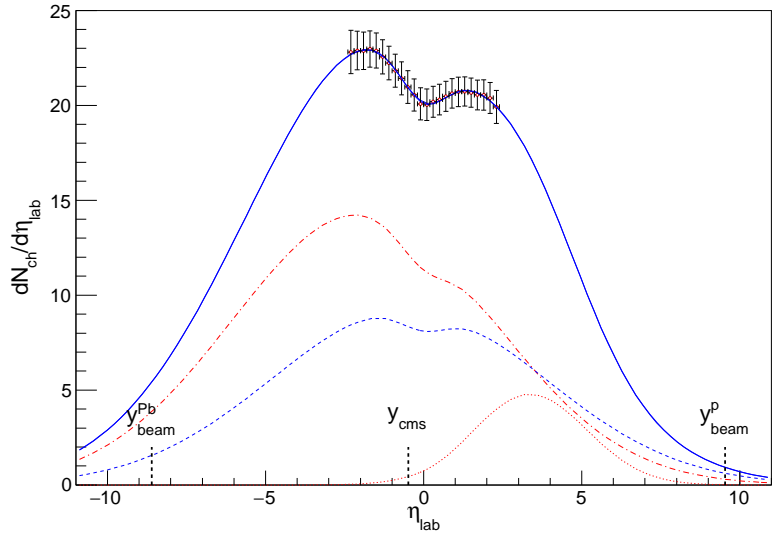}
\caption{\label{fig2}(Color online) The RDM pseudorapidity density distribution function for produced charged hadrons in minimum-bias pPb collisions
at the LHC energy of  $\sqrt{s_\text{NN}}=8.16$ TeV (upper solid curve) has been optimized with respect to the CMS data \cite{cms18}\, as in Fig.\,\ref{fig1}. Here, the underlying distributions in the three-source model are shown: The dashed curve
arises from gluon-gluon collisions, the dot-dashed curve from valence quark-gluon events in the Pb-going region ($y < 0,$ backward), and the dotted curve from valence quark-gluon events in the p-going direction
(fragmentation sources). A corresponding earlier RDM prediction is shown in figure 2 of Ref.\cite {sgw15}{}. }
\end{center}
\end{figure}
\begin{figure}[tph]
\begin{center}
\includegraphics[width=8.0cm]{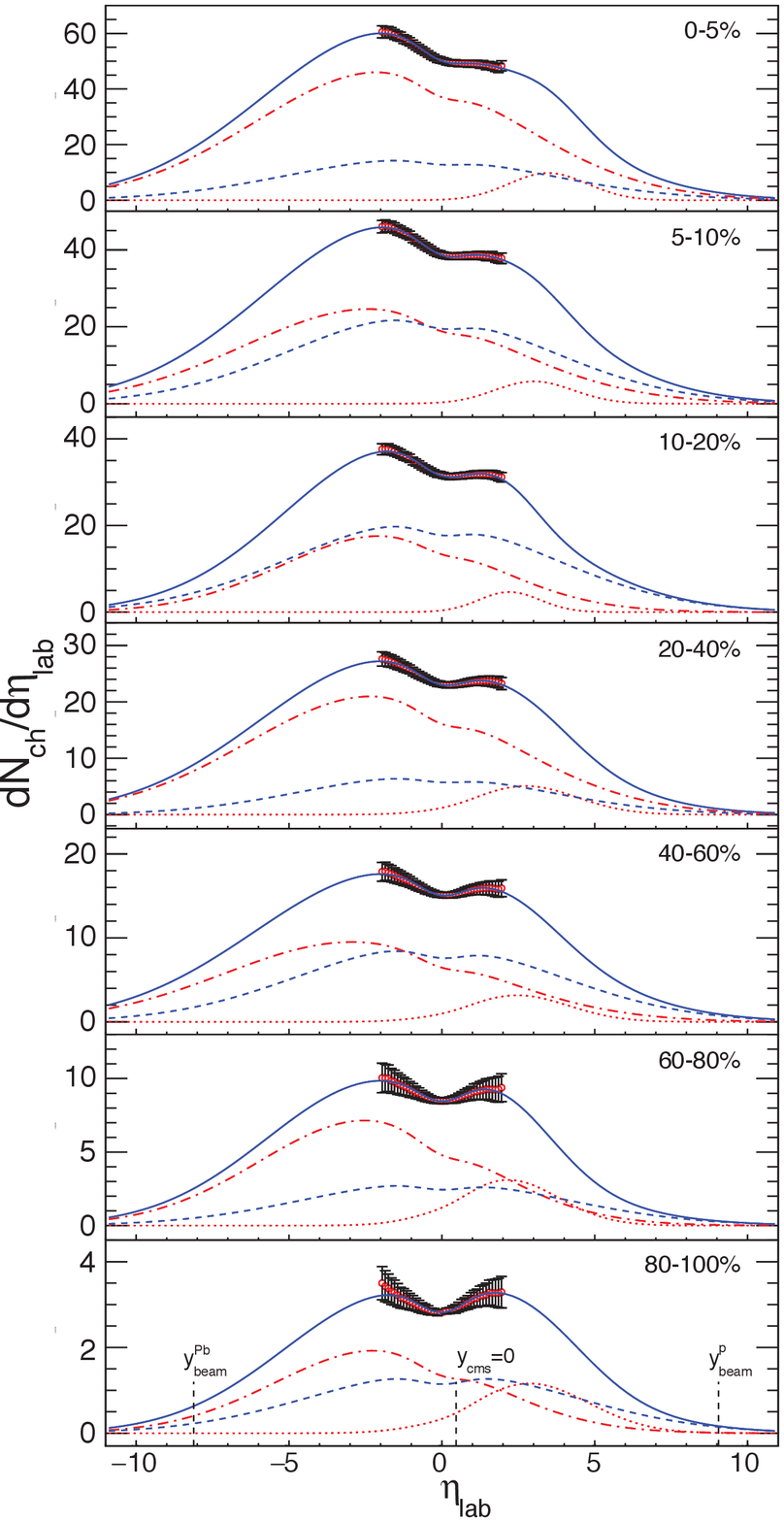}
\caption{\label{fig3}(Color online) The centrality-dependent RDM pseudorapidity distribution functions \cite{gw13} for produced charged hadrons in pPb collisions
at LHC cm energy of 5.02 TeV are adjusted in the mid-rapidity region to the  ALICE data \cite{adam15} through $\chi^2$-minimizations.
The lab frame is shifted by $\Delta y =0.465$ with respect to the cm frame. 
The underlying distributions in the three-source model are also shown, with the dashed curves
arising from gluon-gluon collisions, the dotted curves in the p-going direction ($y > 0,$ forward), and the dot-dashed curves from valence quark-gluon events in the Pb-going region ($y < 0,$ backward; fragmentation sources). Parameters are from table\,\ref{tab1a}.}
\end{center}
\end{figure}
\section{Minimum-bias and centrality-dependent pPb collisions at 5.02 and 8.16 TeV}
Our results for the three-source RDM calculations of produced charged hadrons in minimum-bias pPb collisions at  $\sqrt{s_\text{NN}}=5.02$ TeV and 8.16 TeV are shown in
figure~\ref{fig1} and compared with corresponding CMS data \cite{cms18}. Here the higher energy is the maximum that can currently be achieved for pPb at the LHC, because $\sqrt{s_\text{NN}}= \sqrt{Z_1Z_2/(A_1A_2)}\times 2p_\text{p}$ = 8.16 TeV for $p_\text{p}$ = 6.5 TeV/$c$.
A comparison with a calculation for dAu at 200 GeV done in analogy to earlier results from Ref.\cite{wobi06}\,, and PHOBOS data \cite{bb04} is also shown. 

The distributions fall off less rapidly in the direction of the heavier particle (Au, Pb) as compared to the direction of the light one (p, d). Subdistributions are shown for the dAu system, RDM-parameters are given in table\,\ref{tab1}. Here the $\chi^2$-values are given per number of degrees of freedom (ndf), which is the number of data points minus the number of free parameters of the model. The result for 8.16 TeV agrees quite well
with our RDM-prediction shown in figure 2 of Ref.\cite{sgw15} when the exchange of forward and backward direction is taken into account.

The fragmentation sources are clearly indispensable for an understanding of the data: A thermal-model approach with a single source
that is modified in pseudorapidity space through the Jacobian does not reproduce the very asymmetric data. A thermal model
with several sources such as Ref.\cite {liu04} can reproduce the data, but does not account for the time-dependent nonequilibrium evolution of the partial distributions,
it rather postulates thermal sources at different locations in rapidity space. Also, more than three sources (four in Ref.\cite {liu04}) are difficult to motivate, whereas the two fragmentation sources in the RDM are microscopically mainly due to valence quark-gluon interactions, and the third source is essentially due to low-$x$ gluon interactions.  

Obviously, triple-Gaussian fits in $\eta$-space
as performed in Ref.\cite{bb04} without any consideration of the Jacobian can also be used in $\chi^2$-minimizations to fit the data, but -- different from a phenomenological model -- these have no underlying physical meaning, and do not provide predictions.

Figure\,\ref{fig2}  shows again the result for 8.16 TeV pPb compared to CMS data \cite{cms18}, but now with the corresponding subdistributions. At this higher energy, substantially more charged particles $(N_\text{ch}^\text{tot} = 270)$ are produced as compared to the lower energies
(211 charged hadrons in 5.02 TeV pPb).


In figure~\ref{fig3} with parameters from table\,\ref{tab1a}, we display the centrality dependence of produced charged hadrons in 5.02 TeV pPb collisions at seven centralities, and the
resulting RDM-distributions with subdistributions. The ALICE data are from Ref.\cite{adam15}{}. Towards more central collisions, the midrapidity sources are more important as compared to the fragmentation sources, and the total distribution function becomes progressively more asymmetric. In particular, the slope on the p-going side is steeper than the one on the Pb-going side, as has already been evident from the minimum-bias results. To actually test this model prediction in pPb collisions, data in a larger pseudorapidity range are needed.
\begin{table}[h]
\tbl{RDM-parameters for minimum-bias charged-hadron production in dAu at 0.2 TeV, and in pPb at 5.02 TeV and 8.16 TeV in the lab system ($y_\text{beam}^\text{cm}$ is in the cms).
 The FWHMs of the three sources at the freeze-out time are $\Gamma_k$, the corresponding charged-particle contents are  $N_\mathrm{ch}^k$. The Jacobian scale is $q=m_\pi/\langle p_\perp^\text{eff}\rangle$, see text, and ndf is the number of degrees of freedom.}
 {\begin{tabular}{@{}ccccccccccccc@{}} \toprule
$\sqrt{s_\text{NN}}$~(TeV)&$y_\text{beam}^\text{cm}$&$\langle y_{1} \rangle$&$\langle y_{2} \rangle$& $\Gamma_{1}$ &$\Gamma_{2}$& 
$\Gamma_\text{gg}$&$N_\mathrm{ch}^{1}$&$N_\mathrm{ch}^{2}$&$N_\mathrm{ch}^\text{gg}$&$q$&ndf&$\chi^2/$ndf\\
\colrule
0.20&$\pm5.362$&$2.20$&$-2.90$&$3.40$&$5.90$&$4.73$&$17$&$53$&$22$&$0.45$&$47$&$0.16$\\ 
5.02&$\pm8.586$&$3.01$&$-1.50$&$3.57$&$8.00$&$10.57$&$15$&$105$&$91$&$0.55$&$17$&$0.06$\\ 
8.16&$\pm9.071$&$3.36$&$-1.92$&$4.24$&$9.70$&$10.11$&$22$&$148$&$100$&$0.55$&$17$&$0.01$\\  \botrule
\end{tabular}\label{tab1} }
\end{table}

From Figs.\,\ref{fig2} and \ref{fig3} it is obvious that the tails of the pseudorapidity distributions in pPb and PbPb collisions extend beyond the values of the beam rapidities. This effect had been discussed in Ref. \cite{gw15}\,:  Most of the produced charged hadrons are pions, and the limit $\eta\approx y$ at small transverse momenta (very forward angles) is reached for charged hadrons at larger values of $\eta$ than for protons (net protons determine the value of the beam rapidity). Hence, the $dN/d\eta$ distribution for charged hadrons which are mostly pions can easily extend beyond $y_\text{beam}$. This was shown experimentally in AuAu collisions at RHIC energies by the PHOBOS collaboration \cite{alv11}. At LHC energies, a corresponding measurement is not yet possible. 

The central and Pb-like sources in pPb collisions at 5.02 and 8.16 TeV have a large overlap in $\eta$-space. This particular outcome of the $\chi^2$-minimization is not too surprising because the backward source contains the dominant number of nucleons giving rise to the fragmentation sources, with the forward source emerging from one nucleon initially. The equilibrium value of the rapidity is thus on the Pb-going side, and the center of the midrapidity source is shifted accordingly towards this side. 
The situation is very different in PbPb collisions (Fig.\,\ref{fig4}) with equal particle content in the forward and backward sources, and the equilibrium value of the rapidity being $y_\text{eq} = 0$.
\begin{table}[h]
\tbl{RDM-parameters for centrality-dependent charged-hadron production pPb at 5.02 TeV in the lab system.
 The FWHMs of the three sources at the freeze-out time are $\Gamma_k$, the corresponding charged-particle contents are  $N_\mathrm{ch}^k$. The Jacobian scale is $q=m_\pi/\langle p_\perp^\text{eff}\rangle$, see text. The degrees of freedom are $\text{ndf} = 33$.}
{\begin{tabular}{@{}cccccccccccc@{}} \toprule
Centrality\,(\%)&$\langle y_{1} \rangle$&$\langle y_{2} \rangle$&$y_{eq}$& $\Gamma_{1}$ &$\Gamma_{2}$& 
$\Gamma_\text{gg}$&$N_\mathrm{ch}^{1}$&$N_\mathrm{ch}^{2}$&$N_\mathrm{ch}^\text{gg}$&$q$&$\chi^2$/ndf\\
\colrule
0--5 & $3.474$ & $-1.854$ & $-0.780$ & $3.02$ & $10.00$ & $10.00$ & $32$ & $495$ & $158$ & $0.55$ & $0.04$ \\
5--10 & $3.000$ & $-2.240$ & $-0.726$ & $3.04$ & $10.00$ & $10.00$ & $19$ & $264$ & $241$ & $0.56$ & $0.03$ \\
10--20 & $2.190$ & $-1.845$ & $-0.682$ & $2.38$ & $8.00$ & $10.00$ & $12$ & $151$ & $220$ & $0.57$ & $0.08$\\
20--40 & $2.724$ & $-2.112$ & $-0.589$ & $3.99$ & $9.973$ & $9.40$ & $22$ & $224$ & $67$ & $0.59$ & $0.03$ \\
40--60 & $2.433$ & $-2.854$ & $-0.430$ & $4.03$ & $10.00$ & $10.00$ & $14$ & $101$ & $96$ & $0.62$ & $0.03$ \\
60--80 & $2.107$ & $-2.392$ & $-0.228$ & $4.00$ & $8.42$ & $9.95$ & $13$ &$65$ & $31$ & $0.67$ & $0.03$ \\
80-100 & $2.840$ & $-1.996$ & $-0.001$ &  $4.64$ & $8.12$ & $10.00$ & $6$ & $17$ & $15$ & $0.73$ & $0.13$ \\  \botrule
\end{tabular}\label{tab1a} }
\end{table}

\section{Centrality-dependent PbPb collisions at 5.02 TeV}
At the LHC, PbPb collisions have meanwhile also been measured at a center-of-mass energy of 5.02 TeV in the nucleon-nucleon system, so that these can be directly compared
with pPb. Due to the system's symmetry, the linear relativistic diffusion model has only five parameter for the three subdistributions in rapidity space, as was detailed in section 2: the mean peak positions $\langle y_1\rangle = -\langle y_2\rangle$ (or equivalently, the ratio of freeze-out time and rapidity relaxation time), the variances $\sigma^2$, or the corresponding FWHMs $\Gamma=\sqrt{8\ln2}\,\sigma$ of fragmentation- and midrapidity distributions, and the associated particle numbers.
\begin{figure}[tph]
\begin{center}
\includegraphics[width=7.8cm]{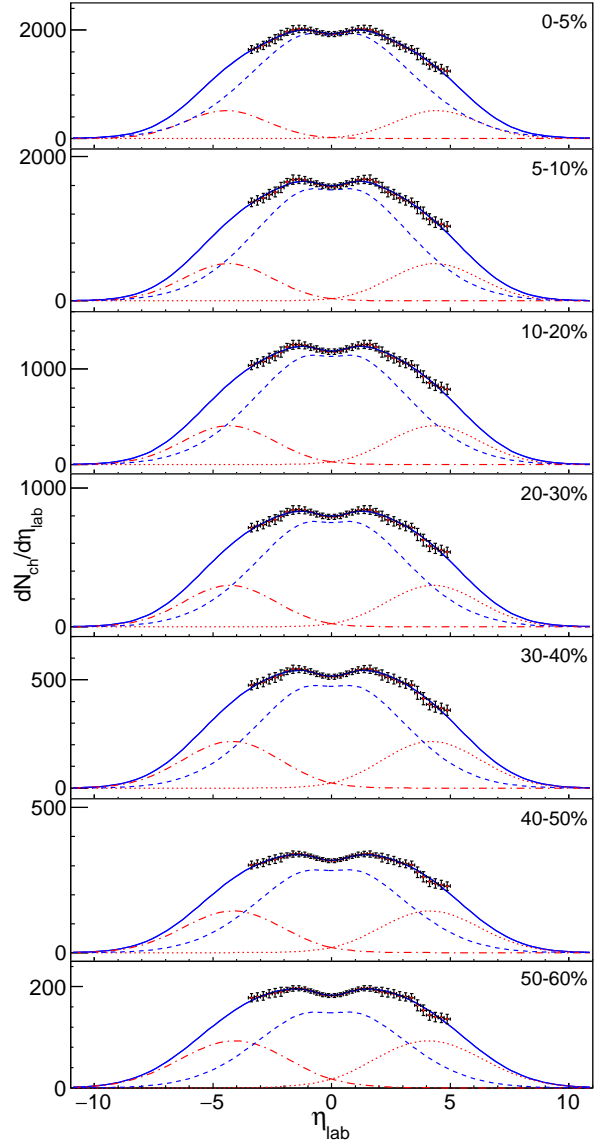}
\caption{\label{fig4}(Color online) The centrality-dependent RDM pseudorapidity distribution functions \cite{gw13} for produced charged hadrons in PbPb collisions
at the LHC energy of 5.02 TeV are adjusted in the mid-rapidity region to the  ALICE data \cite{alice17} through $\chi^2$-minimizations.
The underlying distributions in the three-source model are also shown, with the dashed curves
arising from gluon-gluon collisions, the dotted curves from valence quark-gluon events in the forward-going region ($y > 0$), and the dot-dashed curves in the backward direction
(fragmentation sources). The calculations employ the limiting fragmentation hypothesis with respect to the 0.2 TeV AuAu data
\cite{bea02,alv11}\,, see text.}
\end{center}
\end{figure}

\begin{table}[h]
\tbl{RDM-parameters for charged-hadron production in 5.02 TeV PbPb with $y_{\text{beam}} = \pm\,8.586$ at seven centralities. The FWHM of the sources at the freeze-out time is
$\Gamma_k$, the corresponding charged-particle content $N_\mathrm{ch}^k$. The $\chi^2$/ndf-values refer to the number of degrees of freedom, ndf. The last column gives the experimental midrapidity values from ALICE \cite{alice17}\,.}
{\begin{tabular}{@{}cccccccc@{}} \toprule
Centrality(\,\%)&$\langle y_{1,2} \rangle$& $\Gamma_{1,2}$ &
$\Gamma_\text{gg}$&$N_\mathrm{ch}^{1+2}$&$N_\mathrm{ch}^\text{gg}$&$\chi^2$/ndf&$\frac{\mathrm{d}N}{\mathrm{d}\eta}|_{\eta \simeq 0}$\\
\colrule
0--5 &$\pm 4.44$&$4.02$&$7.38$&$16896$&$21318$&$0.11$&$1929\pm 46$\\
5--10 &$\pm 4.36$&$4.40$&$6.95$&$12815$&$17657$&$0.17$&$1583\pm 37$\\
10--20 &$\pm 4.32$&$4.49$&$6.94$&$9450$&$13347$&$0.15$&$1181\pm 28$\\
20--30 &$\pm 4.28$&$4.54$&$6.85$&$6176$&$9072$&$0.13$&$792.4\pm 18$\\ 
30--40 &$\pm 4.21$&$4.82$&$6.75$&$3806$&$6027$&$0.13$&$514.7\pm 11$\\ 
40--50 &$\pm 4.15$&$4.90$&$6.74$&$2292$&$3796$&$0.12$&$317.5\pm 7$\\ 
50--60 &$\pm 4.08$&$5.41$&$6.71$&$1197$&$2270$&$0.13$&$182.5\pm 4$\\ \botrule
\end{tabular}\label{tab3} }
\end{table}

The resulting pseudorapidity distributions for produced charged hadrons in 5.02 TeV PbPb collisions at seven centralities are displayed in figure~\ref{fig4}. The distribution functions have been determined in $\chi^2$-minimizations of the differences between ALICE data 
\cite{alice17}\, and the analytical RDM-solutions -- including the Jacobian transformation and limiting fragmentation -- by means of the current version of the object-oriented data analysis framework ROOT \cite{brun97}\,. These parameters are summarized in table \ref{tab3} for seven centralities.

As already found in Ref.\cite {rgw12} for the centrality dependence in 2.76 TeV PbPb, the RDM-parameters depend monotonically on the number of participants. We have described in section 2 how to use the limiting fragmentation conjecture also at 5.02 TeV PbPb relative to the 200 GeV AuAu PHOBOS data \cite{bea02,alv11}  in the tails of the distribution functions, and the Jacobian transformation from $y$- to $\eta$-space with adapted values of $\langle p_\perp^\text{eff}\rangle$.

At all centralities, the mid-rapidity gluon-gluon source (dashed curves in figure~\ref{fig4}) has a larger particle content than the sum of the two fragmentation sources, with a relative charged-hadron content rising monotonically from $\sim 56$\,\% in the $0-5\%$ centrality bin to $\sim 65$\,\% in the $50-60\%$ centrality bin. The evolution of the particle content in in three subdistributions with cm energy had already been discussed in Ref.\cite {gw15}\, where analytical formulae for the respective dependencies had been given. These are in line with the present results of the $\chi^2$-minimizations.

\section{Prediction for central PbPb collisions at 39 TeV}
We also make a prediction for central PbPb collisions at the Future Circular Collider (FCC), which is currently discussed in design studies for pp collisions with $\sqrt{s}=100$ TeV,
corresponding to PbPb with $\sqrt{s_\text{NN}}=Z/A\times100\,\text{TeV}=39.42$ TeV and $y_\text{beam}=\pm 10.646$. 

The five RDM parameters have been extrapolated to this energy
according to the analytical formulae given in Ref.\cite {gw15}{}, see table\,\ref{tab3}. Regarding the Jacobian, both the mean mass of the produced charged hadrons, and also the mean transverse momentum
are expected to rise, and it is difficult to predict the energy dependence of their quotient, which essentially determines the Jacobian scale $q$. For the purpose of a rough estimate, we have
therefore taken the Jacobian scale to be unchanged from 5.02 TeV.

\begin{figure}[tph]
\begin{center}
\includegraphics[width=12.4cm]{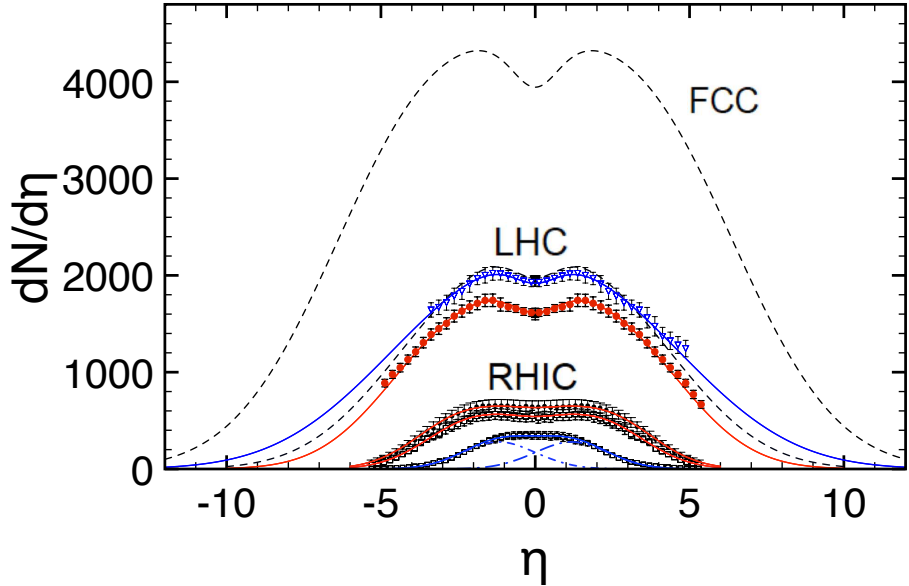}
\caption{\label{fig5}(Color online) The RDM pseudorapidity distribution functions for produced charged hadrons in central (0-5\%) AuAu (RHIC) and PbPb (LHC) collisions
at cm energies of 19.6 GeV, 130 GeV, 200 GeV, 2.76 TeV and 5.02 TeV (bottom up) are optimized in $\chi^2$-fits with respect to the PHOBOS \cite{bb03,alv11} (bottom) and ALICE \cite{abb13,alice17} (middle) data, with parameters from Ref.\cite {gw13,gw16}\, and table \ref{tab3}.
The middle dashed curve is a prediction at 5.02 TeV within the three-source model from Ref.\cite {gw16}\,, the solid curve the result of a $\chi^2$-minimization including limiting fragmentation scaling.
The upper dashed curve is a RDM-prediction for PbPb at the FCC energy of $\sqrt{s_\text{NN}}= 39.24$ TeV with $y_\text{beam}=\pm 10.646$, parameters are from table\,\ref{tab3}.}
\end{center}
\end{figure}

The resulting prediction at the projected FCC energy is shown together with central AuAu and PbPb results at RHIC and LHC energies in figure~\ref{fig5}. The RDM calculation produces a higher midrapidity value $(dN/d\eta \simeq 3944)$ when compared to the result of the empirical extrapolation of midrapidity values as performed in Ref.\cite{gw15} (with $s_0=1$ TeV) 
\begin{equation}
\frac{dN_\text{ch}}{d\eta}\Bigr |_{\eta\simeq 0}=1.15\times10^3(\sqrt{s_\text{NN}}/s_0)^{0.33} \simeq 3860.
\label{fcc}
\end{equation}
It remains to be seen which of the two values for the midrapidity yield of produced charged hadrons is more reliable.
\begin{table}[h]
\tbl{RDM-parameters for charged-hadron production extrapolated to 39.423 TeV PbPb with $y_{\text{beam}} = \pm\,10.646$ at 0-5\% centrality. $\Gamma$ is the FWHM of the sources at the freeze-out time, $N_\mathrm{ch}$ the corresponding charged-particle content using the extrapolation formulae of Ref.\cite{gw15}\,. The last column gives the predicted midrapidity value according to  Ref.\cite{gw15}\,.}
{\begin{tabular}{@{}ccccccc@{}} \toprule
Centrality&$\langle y_{1,2} \rangle$& $\Gamma_{1,2}$ &
$\Gamma_\text{gg}$&$N_\mathrm{ch}^{1+2}$&$N_\mathrm{ch}^\text{gg}$&$\frac{\mathrm{d}N}{\mathrm{d}\eta}|_{\eta \simeq 0}$\\
\colrule
0--5\,\%&$\pm 4.38$&7.0&9.2&14354&30917&$3860$\\ \botrule
\end{tabular}\label{tab3} }
\end{table}

\section{Conclusions}
We have analyzed pPb and PbPb collisions at LHC cm energies of 5.02 TeV, and pPb collisions at 8.16 TeV using the linear version of the nonequilibrium-statistical relativistic diffusion model, and determined the model parameters in fits to the available data. The asymmetric pPb system is particularly sensitive to the interplay of the three subdistributions in the RDM.

At all centralities, the mid-rapidity source has the largest particle content, but the fragmentation sources are necessary for a detailed modeling of the centrality-dependent shape of the total pseudorapidity distribution functions of produced charged hadrons. The incoherent superposition of the three subdistributions together with the effect of the Jacobian transformation from rapidity to pseudorapidity determines the shape of the distributions. For minimum-bias pPb collisions at 8.16 TeV the result is in reasonable agreement with an earlier RDM-prediction \cite{sgw15}.

The shapes of the total distribution functions -- in particular, for asymmetric systems -- show that the system has not reached overall statistical equilibrium: The centers of the fragmentation distributions remain far from the equilibrium values $y_\text{eq}(b)$. Still, the midrapidity results, as well as integrated yields, may be close to thermal-model predictions. 

In symmetric systems such as PbPb, the deviations from equilibrium distributions in $\eta$-space including collective expansion may seem less obvious from the data, but the underlying subdistributions show that these are also far from equilibrium. Of course, local equilibrium in the hydrodynamic sense is  nevertheless achieved very early in the course of the collision, within a time scale of~$0.1\,\text{fm}/c\lesssim \tau_\text{eq}\lesssim 1\, \text{fm}/c$ -- which is mainly due to the very short local equilibration time \cite{gw18} of the dense gluon system that characterizes the early stages of the collision.

With RDM-parameters extrapolated to the projected FCC energy of 39 TeV in PbPb, we have also presented a prediction for the pseudorapidity distribution of produced charged hadrons in central collisions at very high energy, but it will take some time to confront it with experimental results.

\section*{Acknowledgments}
We thank Elif Yildirim for assistance with the numerical calculations in the course of her
Heidelberg BSc thesis, and Silvia Masciocchi for discussions.
\bibliographystyle{epj.bst}
\bibliography{gw_mpla}

\end{document}